\documentclass[aps,prl,reprint,groupedaddress,floatfix]{revtex4-2}
\usepackage{bm}
\usepackage{amsmath}
\usepackage{graphicx}% Include figure files
\usepackage{dcolumn}%
\usepackage{color}
\usepackage{comment}
\usepackage{hyperref}
\begin{document}

%\preprint{}

\title{Persistence in Active Turbulence}


\author{Amal Manoharan}
\affiliation{Department of Physics, Indian Institute of Technology Madras, Chennai - 600036}
\author{Sanjay CP}
\affiliation{International Center for Theoretical Sciences, Bengaluru, 560089}
\author{Ashwin Joy}
\email[]{ashwin@physics.iitm.ac.in}
\affiliation{Department of Physics, Indian Institute of Technology Madras, Chennai - 600036}

%\thanks{}

\date{\today}

\begin{abstract}
  Active fluids such as bacterial swarms, self-propelled colloids, and cell tissues can all display complex spatio-temporal vortices that are reminiscent of inertial turbulence. This emergent behavior despite the overdamped nature of these systems is the hallmark of active turbulence. In this letter, using a generalized hydrodynamic model, we present a study of the persistence problem in active turbulence. We report that the persistence time of passive tracers inside the coherent vortices follows a Weibull probability density whose shape and scale are decided by the strength of activity \textemdash contrary to inertial turbulence that displays power-law statistics in this region. In the turbulent background, the persistence time is exponentially distributed that is remindful of inertial turbulence. Finally we show that the driver of persistence inside the coherent vortices is the temporal decorrelation of the topological field, whereas it is the vortex turnover time in the turbulent background.

\end{abstract}

%\keywords{}

\maketitle
\begin{figure}[]
  \includegraphics[width=\linewidth,keepaspectratio]{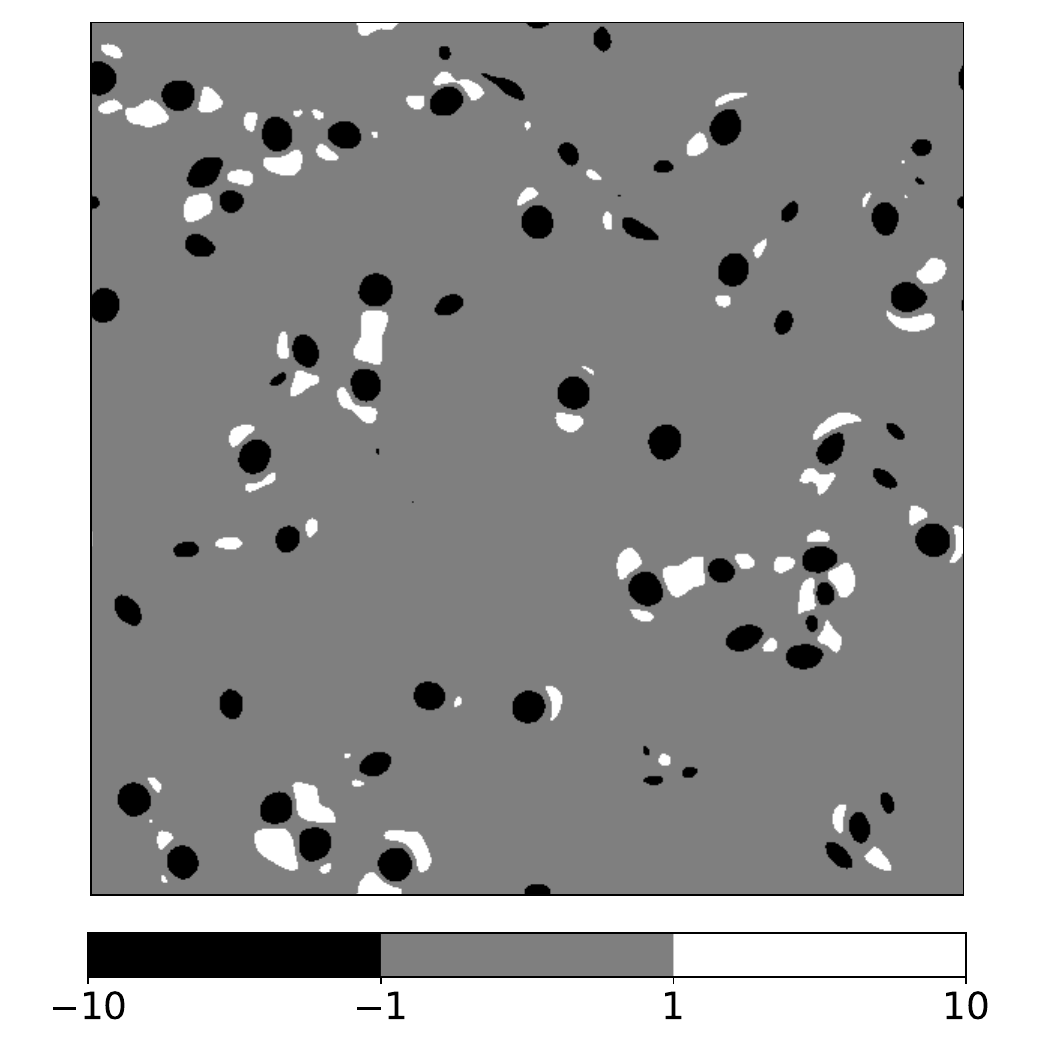}
  \caption{Okubo-Weiss field of the turbulent fluid normalized to its rms value. The snapshot is taken in the steady state. We can clearly identify three topologically distinct regions, namely rotation dominated ($\mathcal{Q} < -1$), deformation dominated ($\mathcal{Q} > 1$) and intermediate ($- 1 \leq \mathcal{Q} \leq 1$) regions.}
  \label{topology_of_fluid}
\end{figure}

Persistence in physical systems concerns with the probability that a local fluctuating field does not change its sign upto a time $t$. From Ising spins \cite{PhysRevLett.75.751}, rough surfaces \cite{PhysRevE.56.2702} and disordered media \cite{PhysRevLett.82.3944}, to  optimization \cite{brown1997optimization}, machine learning \cite{widmer1996learning} and stock markets \cite{PhysRevE.72.051106} \textemdash persistence often contains important information about the evolution history of a complex system. Theoretical investigations have shown that the probability density function of this persistence time $\mathcal{P}(t)$ is non-trivial because the underlying fluctuating field is usually non-Markovian \cite{majumdar1999persistence}. A related and often useful quantity is the mean first passage time distribution of a particle diffusing in a bounded media. Once computed, such distributions can be profitably exploited to quantify and compare structural correlations and dynamical heterogeneity in complex systems \cite{bassolas2021first, li2013mechanisms} \textemdash directly allowing a characterization of energy landscapes in these complex systems \cite{bebon2022first}. While there exists a large body of work on persistence and first-passage problems in many-body systems far from equilibrium \cite{majumdar1999persistence,bray2013persistence,lancaster2019persistence,jose2022first,salcedo2022persistence}, similar investigations in active or ``self-propelled''  systems have been very few. Thus, there is a pressing need to explore persistence and first passage in active systems, especially with models that allow a general pattern of energy injection, transfer and dissipation. In this Letter, we report a careful Lagrangian study of persistence time and its distribution in a model active liquid that can display spatio-temporal vortices that are remindful of classical turbulent flows. To this end, we invoke the Okubo-Weiss criterion \cite{okubo1970horizontal,weiss1991dynamics} to perform a topological partitioning of the active flow field  into rotation dominated, deformation dominated, and intermediate regions \textemdash see Fig. \ref{topology_of_fluid} for a visualization. We show that in the rotation dominated regions, $\mathcal{P}(t)$ is given by the Weibull probability density whose shape and scale are decided by the strength of activity. In the deformation dominated regions, $\mathcal{P}(t)$ is exponential. Both these observations are contrary to the case of inertial turbulence that displays power law statistics in these regions \cite{kadoch2011lagrangian,PhysRevLett.106.054501}. In the intermediate region that forms the turbulent background, $\mathcal{P}(t)$ is exponentially distributed \textemdash beautifully remindful of inertial turbulence. We believe our work is a novel study of the persistence time distributions  in active matter flows that are valid over a wide range of a control parameters, thereby putting a large number of active systems under the purview of work \textemdash from elementary forms of life, like bacterial suspensions to synthetic active matter such as Janus colloids.

\begin{figure}[!ht]
\centering
   \includegraphics[width=1\linewidth]{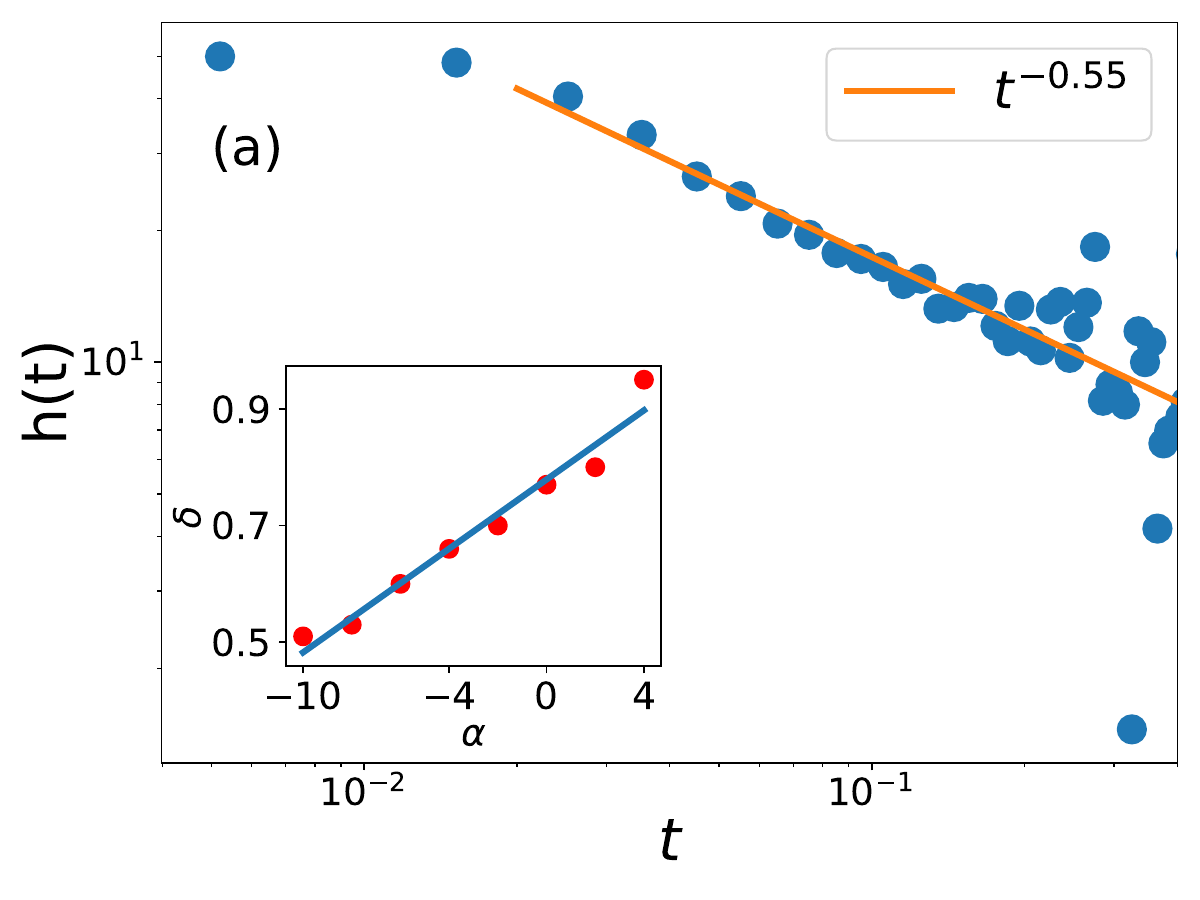}
   \includegraphics[width=1\linewidth]{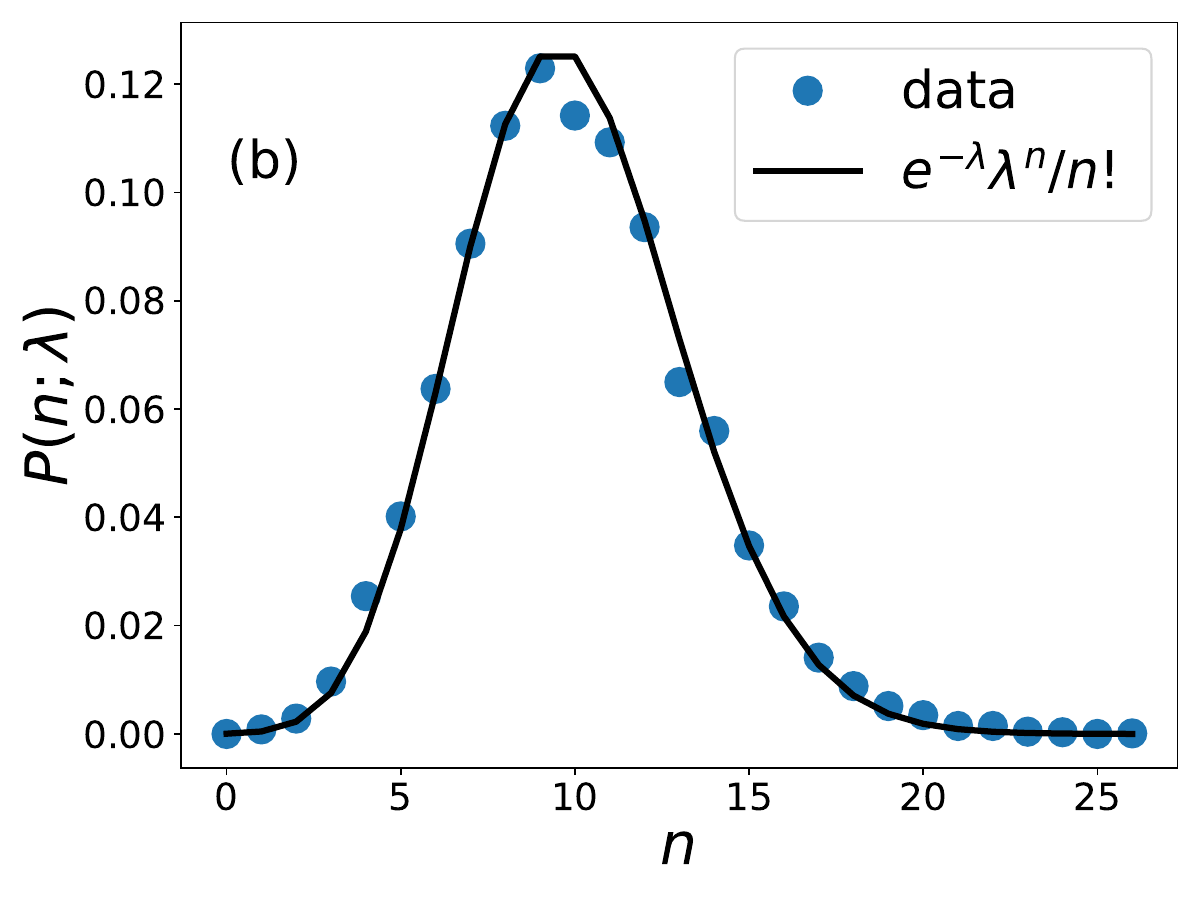}
   \caption{(a) The hazard function (Eq. \ref{hazard-eq} plotted as a function of time for a given $\alpha=-10$. The power law fit verifies that the shape exponent $\delta \approx 0.45$. Inset shows that the Weibull shape parameter $\delta$ is linear in $\alpha$. (b) The probability of $n$ tracers exiting the intermediate region over a time interval of $10 \langle t \rangle$. The flow is driven at activity $\alpha = -10$.}
\label{mechanism}
\end{figure}

\textit{Model and simulation:} We perform direct numerical simulations of a generalized hydrodynamic model that is known to reproduce the flow field of dense bacterial suspensions in laboratory experiments \cite{wensink2012meso,dunkel2013minimal,PhysRevLett.110.228102}. In two dimensions, the incompressible velocity field of this model is governed by
\begin{eqnarray}
  \frac{\partial \bm u}{\partial t} + \lambda_0(\bm u
\cdot\bm\nabla)\bm u &=& -\bm\nabla P - \Gamma_0\bm\nabla^2\bm u - \Gamma_2\bm\nabla^4\bm u-\mu \bm u\nonumber\\
  \bm \nabla \cdot \bm u &=& 0
  \label{active_model}
\end{eqnarray}
where $P$ is the pressure and the non-dimensional parameter $\lambda_{0}$ decides the type of bacteria, meaning they are either \textit{pusher} ($\lambda_{0} > 0$) or a \textit{puller}  ($\lambda_{0} < 0$) type.  Keeping $\Gamma_{0,2} > 0$ mimics energy injection into the active fluid via instabilities. The scalar field $\mu = \alpha + \beta |\bm u|^2$ depends on the local velocity $\bm u$, and was first introduced by Toner and Tu  to model the ''flocking'' behavior in self-propelled rod-like objects
\cite{toner2005hydrodynamics,toner1998flocks}.  The parameter $\alpha$, henceforth referred to as the Ekman friction, acts at intermediate scales and can either lead to a damping of energy when $\alpha > 0$ or an injection of energy when $\alpha <0$.  Former leads the fluid to an isotropic equilibrium and the latter yields  a globally ordered polar state with mean velocity $\sqrt{|\alpha|/\beta}$. We normalize all distances to a characteristic length $\sigma_0 = 5\pi\sqrt{2\Gamma_2 /\Gamma_0}$ and all times to $t_0 = 5\pi\sqrt{2}\Gamma_2/\Gamma_0^2$. In terms of these reduced units, we fix the values of the model parameters as $ \Gamma_0 = (5\pi\sqrt{2})^{-1}, \Gamma_2 = (5\pi\sqrt{2})^{-3}$ and $\beta = 0.5$, in order to remain consistent with
literature. Equation (\ref{active_model}) is then numerically solved using a pseudo-spectral approach over a square grid of $512^{2}$ points in a doubly periodic box of size $2\pi$. For $\alpha < -6$, we use bigger boxes of size upto $10\pi$ and with resolution upto $2048^2$ to avoid forming condensates. We overcome the aliasing errors that arise due to the implementation of discrete Fourier transforms by performing 2/3 and 1/2 dealiasing rules respectively for the quadratic ($(\bm u\cdot\bm\nabla)\bm u$) and cubic (($|\bm u|^2)\bm u$) terms \cite{canuto2012spectral}. Time marching of $\bm u$ is done using Crank-Nicolson scheme with a time step of $\Delta t = 2 \times 10^{-4}$ that is sufficient to maintain numerical stability in the entire range of parameters explored here. 
\begin{figure}[!ht]
 \includegraphics[width=\linewidth,keepaspectratio]{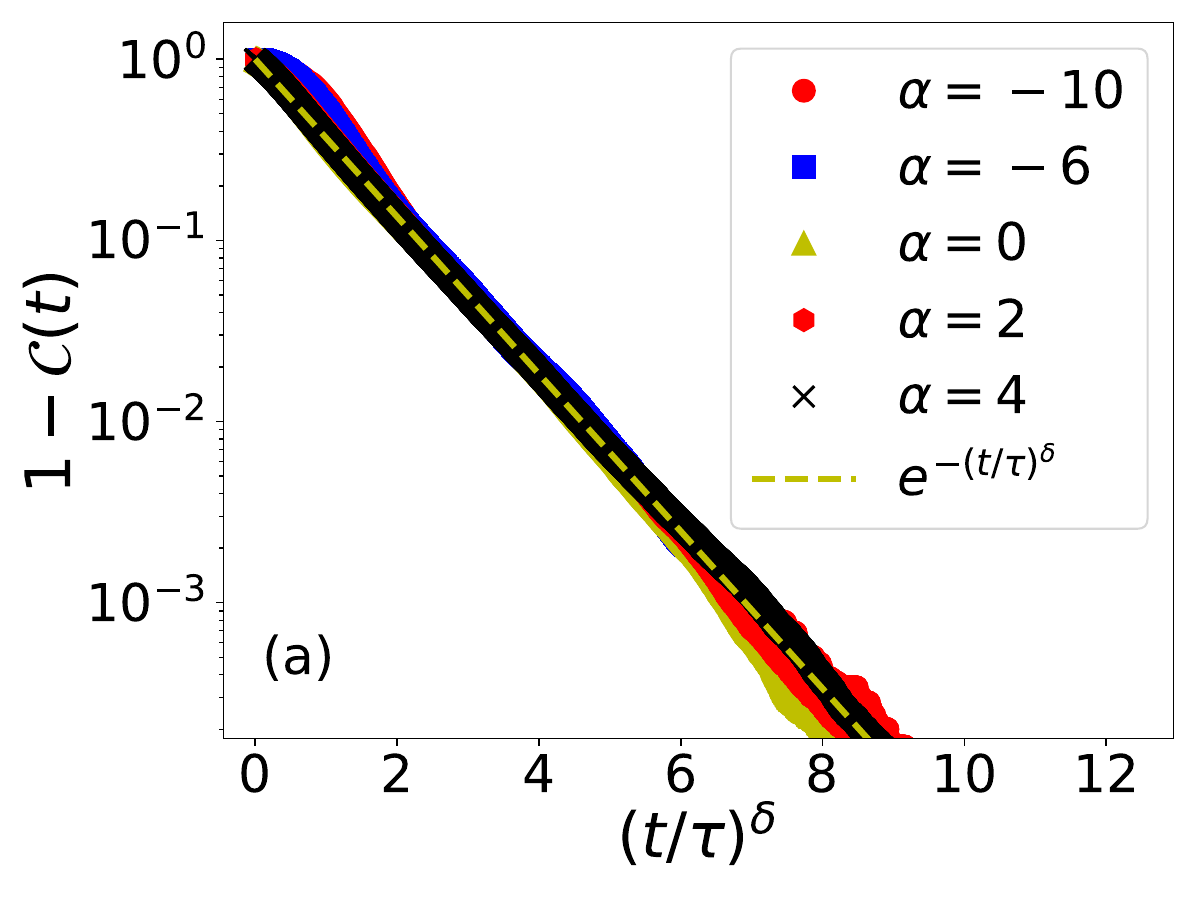}
  \includegraphics[width=\linewidth,keepaspectratio]{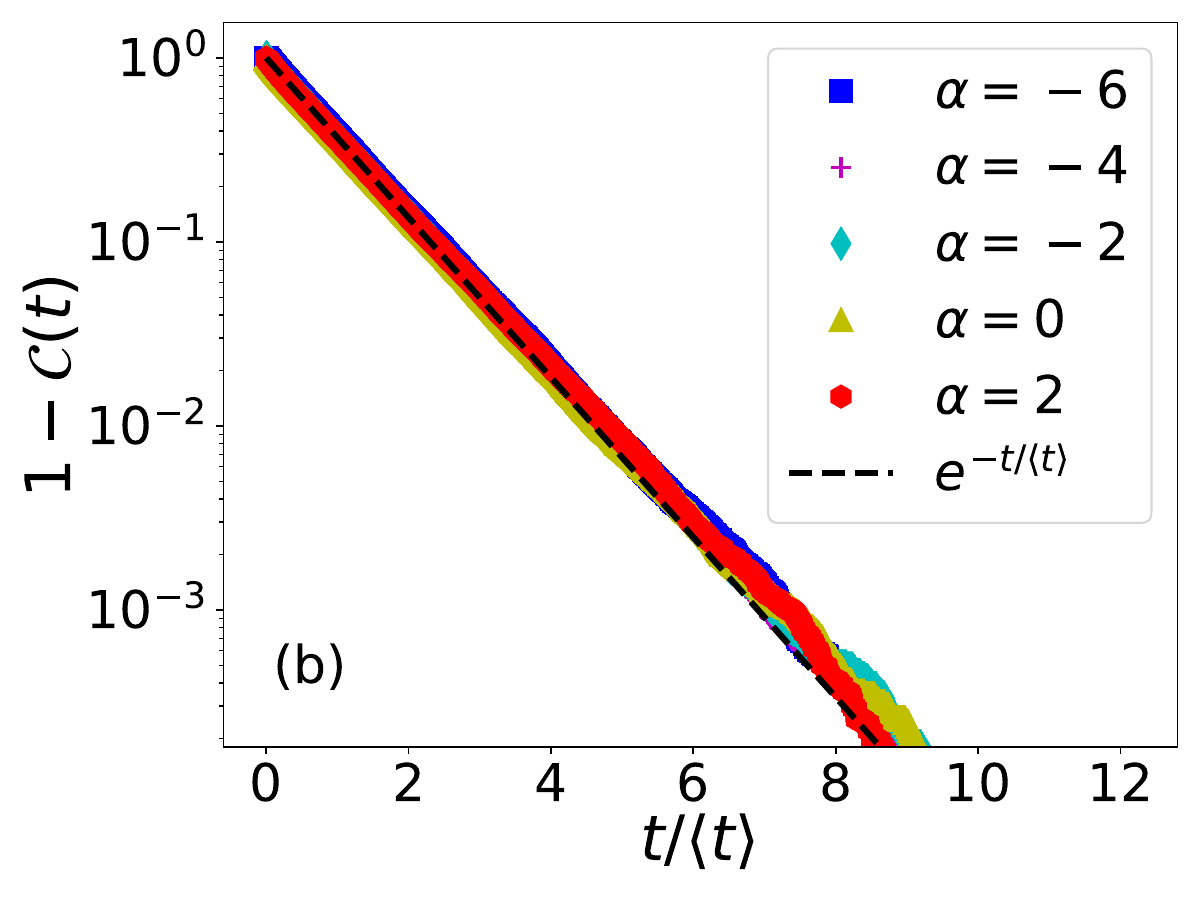}
 \label{vortical}
 \caption{Cumulative distribution function $\mathcal{C}(t)$ of the persistence time in the rotation dominated region (a) and the turbulent background (b). Data collapse shows that the persistence probability $\mathcal{P}(t)$ is a Weibull distribution in the former and an exponential in the latter. The deformation dominated region also shows exponential persistence probability (not shown here).}
 \label{CDF}
\end{figure}
To get Lagrangian statistics, we disperse a distribution of $N$ tracers that follow the dynamics 
\begin{equation}
  \frac{\text{d}\bm x_i(t)}{\text{d}t} = \bm{u}(\bm x_i(t))
  \label{eq-tracer}
\end{equation}
where $\bm x_i(t)$ and $\bm{u}(\bm x_i(t))$ are respectively the tracer location and its velocity. We use a cubic spline interpolation to project Eulerian quantities at any tracer location. As a thumb rule, we disperse these tracers and record their statistics only after the fluid attains a turbulent steady state. To compute the persistence time, we first introduce a fluctuating field that naturally lends a topological characterization of the flow field \cite{kadoch2011lagrangian}. We do this by realizing that the velocity gradient  of an incompressible fluid is a sum of rotation  and deformation tensors
\begin{equation}
  \bm \nabla \bm u = \frac{1}{2}\begin{bmatrix}
    0 & -\omega\\
    \omega & 0
  \end{bmatrix} + \frac{1}{2}\begin{bmatrix}
    \sigma_n & \sigma_s\\
    \sigma_s & -\sigma_n
  \end{bmatrix} 
  \label{gradu}
\end{equation}
with the eigen values $\gamma^2 = (\sigma_n^2 + \sigma_s^2 - \omega^2)/4 = \mathcal{Q}$. Here $\sigma_n = \partial_x u_x - \partial_y u_y$ and $\sigma_s = \partial_x u_y + \partial_y u_x$ denote the normal and shear strains respectively, and $\omega = \partial_x u_{y} - \partial_y u_x$ is the fluid vorticity. Normalized to its root mean squared value, the Okubo-Weiss field $\mathcal{Q}$ can now be used to partition the fluid into three topologically distinct regions, namely rotation dominated for $\mathcal{Q} < - 1$, deformation dominated for $\mathcal{Q} > 1 $ and intermediate for $-1 < \mathcal{Q} < 1$. See Fig. \ref{topology_of_fluid} for a visualization of this field in the steady state. It is evident from this figure that the intermediate region accounts for majority of the area fraction of the fluid. We can now conveniently track the Lagrangian persistence time of any tracer initially seeded in one of these three regions. The distribution of $N$ tracers naturally yields a distribution of persistence times that has a probability density $\mathcal{P}(t)$. It is a standard practice to obtain $\mathcal{P}(t)$ from its cumulative distribution function $\mathcal{C}(t) = \int_0^{t}\mathcal{P}(t')\text{d}t'$ as the procedure is immune to binning errors. In Fig. \ref{CDF}a, we plot this distribution function for the topologically distinct regions in the steady state. The reader should immediately notice that in the region where rotation dominates, $\mathcal{P}(t) = (t/\tau)^{\delta-1}(\delta/\tau)e^{-(t/\tau)^\delta}$ is a Weibull distribution function \textemdash quite contrary to the case of inertial turbulence that shows power law statistics in this region \cite{kadoch2011lagrangian,PhysRevLett.106.054501}. The parameters $\delta$ and $\tau$ denote respectively the shape and scale of this stretched distribution that is often seen in systems displaying extreme value statistics. We observe that $\delta < 1$ throughout the range of our simulations, essentially implying that the hazard function (see later) is a monotonically decreasing function of time. To verify the statistics, we note that the Weibull distribution, at its core, is defined by a simple conditional density, given that the event in question has not occurred yet. Put simply, this is expressed as a hazard function
\begin{equation}
  h(t) = -\dfrac{\text{d}}{\text{d}t} \text{ln} (1-\mathcal{C}(t)) = \dfrac{\delta}{\tau}\bigg(\frac{t}{\tau}\bigg)^{\delta-1}
  \label{hazard-eq}
\end{equation}
In Fig. \ref{mechanism}a, we show a typical plot of the hazard function at a sufficiently high activity of $\alpha=-10$. The power law fit verifies the shape exponent $\delta \approx 0.45$ at this activity. In the inset we show the shape exponent $\delta$ that is linear in the activity strength $\alpha$. Notice that the hazard function decreases with time indicating a continuously falling failure rate. This happens because the tracers seeded near the edge of the vortex are likely to exit sooner than the ones seeded in the core. This is analogous to population dynamics with significant infant mortality leading to the failure rate decreasing over time as weaker infants are removed from the population. We now turn our attention to the intermediate region characterized by $-1 \leq \mathcal{Q} \leq 1$. This is a turbulent background where the density  $\mathcal{P}(t) = e^{-t/\langle t \rangle}/\langle t \rangle$ is an exponential distribution  that is remindful of the inertial turbulence, see Fig. \ref{CDF}b. To verify this, we realize that when the waiting time is exponentially distributed with a mean $\langle t\rangle$, the probability of $n$ tracers exiting the intermediate region over a time interval $\lambda \langle t \rangle$ must be the Poisson probability distribution 
\begin{equation}
  P(n;\lambda) = e^{-\lambda} \frac{\lambda^n}{n!}
  \label{Poisson-eq}
\end{equation}
Indeed our data on event probability fits nicely to the Poisson distribution, see Fig. \ref{mechanism}b. We are thus  drawn to the fact that the exit of tracers from the intermediate region is a memoryless stochastic point process \textemdash a similarity with inertial turbulence that is worth noting. The density $\mathcal{P}(t)$ in the deformation dominated region is also exponential (not shown here) in contrast to the inertial turbulence that exhibits power law scaling \cite{kadoch2011lagrangian}. The curious reader might wonder if this persistence is driven by an intrinsic time scale of the turbulent fluid. This is discussed next.

\begin{figure}[!ht]
  \includegraphics[width=\linewidth]{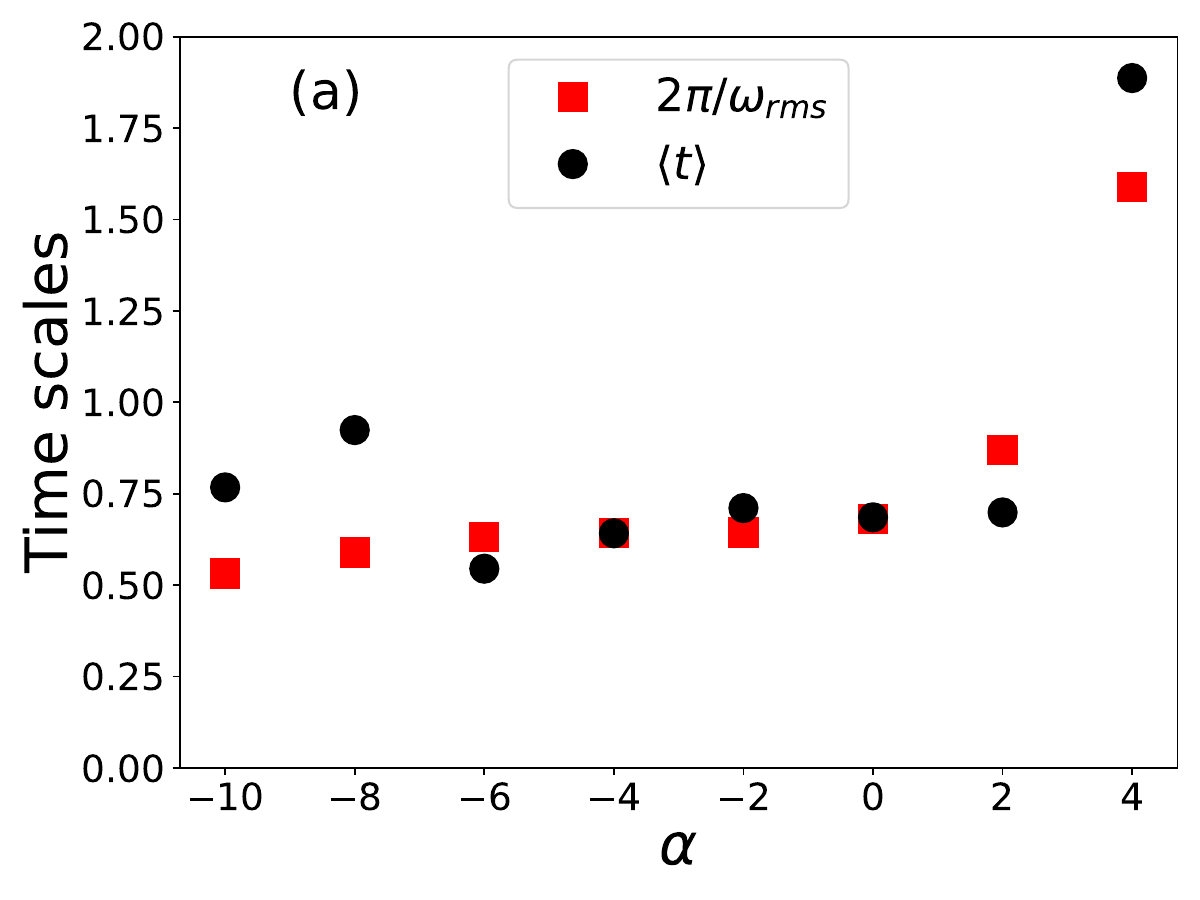}
  \includegraphics[width=\linewidth]{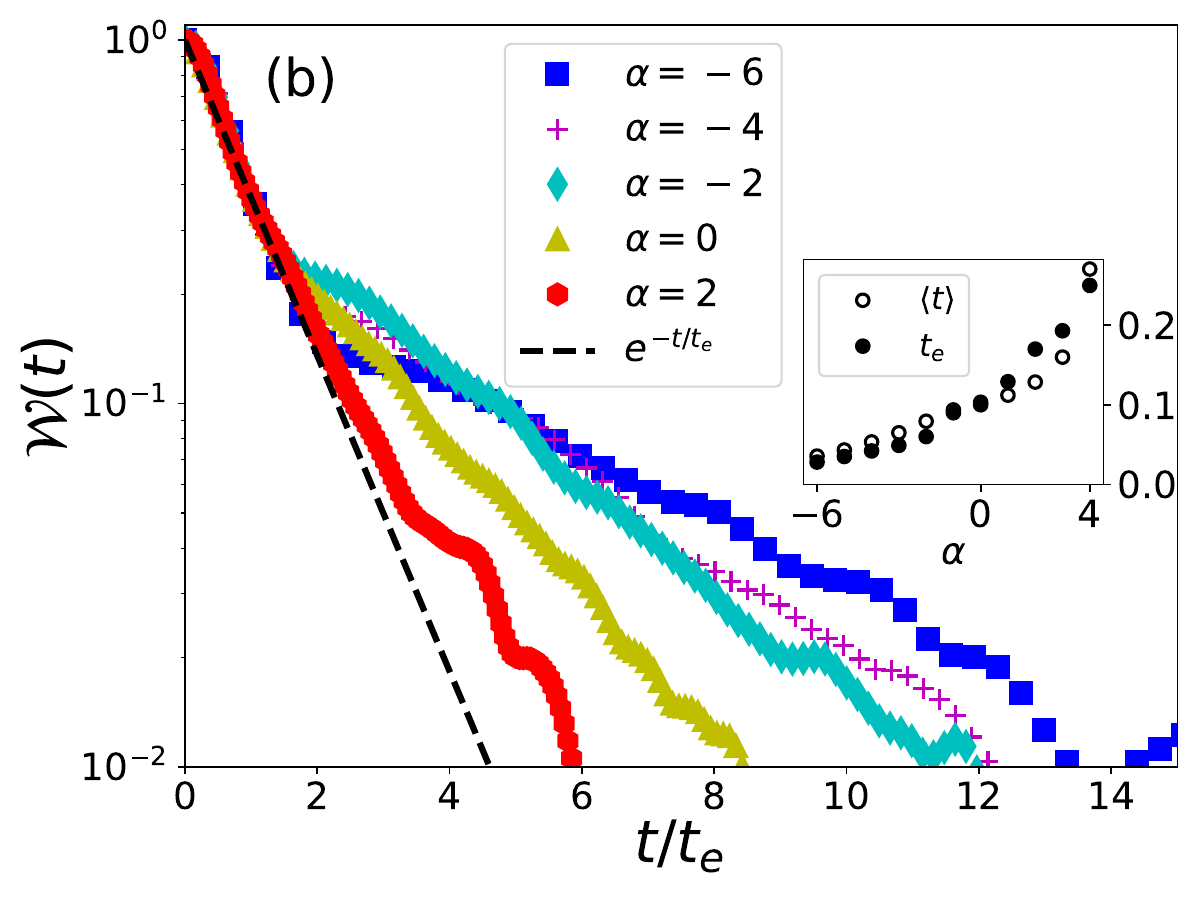}
  \caption{(a) Comparison of the mean persistence time $\langle t \rangle$ with the vortex turn over time  $2\pi/\omega_{\text{rms}}$ in the intermediate region. The two time scales agree well over a wide range of activity, clearly indicating that the vortex rotation rate governs the persistence time \textemdash faster rotation leads to smaller persistence time. (b) Lagrangian autocorrelation of the Okubo-Weiss field. The initial decay is clearly exponential as seen from the data collapse. The e-folding time scale of the decay $t_e$ agrees very well with the mean persistence time $\langle t \rangle$ throughout the range of activity explored (inset).}
  \label{driver}
\end{figure}

In the intermediate region, the root mean squared vorticity $\omega_{\text{rms}}$ can be used to compute a characteristic turnover time as $2\pi/\omega_{\text{rms}}$. We find that this timescale agrees well with the mean persistence time throughout the range of activity explored in our work, see Fig. \ref{driver}a. It is therefore plausible to think of vorticity wandering as the driver of persistence in this region. In the rotation dominated region, we turn our attention to the time auto-correlation of the Lagrangian Okubo-Weiss field $\mathcal{W}(t) = \langle \mathcal{Q}(t)\mathcal{Q}(0) \rangle/\langle \mathcal{Q}(0)^2\rangle$, where $\langle \dots \rangle$ indicates a joint average over the tracers as well as the initial conditions. This is plotted in Fig. \ref{driver}b for various levels of activity. Clearly there is an initial exponential decay that allows a data collapse upto a time scale of the order $t_e$. This is followed by a stretched exponential at late times $t \gg t_e$ that decays faster with increasing $\alpha$ resulting in  ``shorter'' trapping times. The e-folding time scale $t_e$ agrees very well with the mean persistence time which progressively increases with $\alpha$ \textemdash see inset Fig. \ref{driver}b. We also note that $\mathcal{W}(t)$ clearly retains memory in contradistinction to the turbulent background where it is memoryless. We thus conclude here that the driver of persistence in the vortical region is the relaxation of the Okubo-Weiss field  in time.

To our knowledge, this paper constitute a novel study of the persistence in dense swarms of active matter. We do this by a topological partitioning  of the flow field that lends a natural characterization of the flow in terms of regions dominated by rotation, deformation and background turbulence. By observing passive tracers that just go with the flow, we find that their persistence time inside the coherent vortices follows a Weibull distribution whose shape and scale decided by the level of activity. In the turbulent background outside of these vortices, persistence time is exponentially distributed, beautifully remindful of inertial turbulence. We also show that driver of persistence inside the coherent vortices is the temporal decorrelation of the topological field, whereas it is the vortex turnover time in the turbulent background. We believe our findings could be relevant to experiments targeting dense bacterial swarms.

\begin{acknowledgments}
 We thank Abhijit Sen, Sumesh Thampi and Ethayaraja Mani for discussions and comments on the manuscript. Support from the core research grant CRG/2020/001980 from SERB, Government of India, is gratefully acknowledged.
\end{acknowledgments}

\bibliography{manuscript}

\begin{thebibliography}{24}%
\makeatletter
\providecommand \@ifxundefined [1]{%
 \@ifx{#1\undefined}
}%
\providecommand \@ifnum [1]{%
 \ifnum #1\expandafter \@firstoftwo
 \else \expandafter \@secondoftwo
 \fi
}%
\providecommand \@ifx [1]{%
 \ifx #1\expandafter \@firstoftwo
 \else \expandafter \@secondoftwo
 \fi
}%
\providecommand \natexlab [1]{#1}%
\providecommand \enquote  [1]{``#1''}%
\providecommand \bibnamefont  [1]{#1}%
\providecommand \bibfnamefont [1]{#1}%
\providecommand \citenamefont [1]{#1}%
\providecommand \href@noop [0]{\@secondoftwo}%
\providecommand \href [0]{\begingroup \@sanitize@url \@href}%
\providecommand \@href[1]{\@@startlink{#1}\@@href}%
\providecommand \@@href[1]{\endgroup#1\@@endlink}%
\providecommand \@sanitize@url [0]{\catcode `\\12\catcode `\$12\catcode
  `\&12\catcode `\#12\catcode `\^12\catcode `\_12\catcode `\%12\relax}%
\providecommand \@@startlink[1]{}%
\providecommand \@@endlink[0]{}%
\providecommand \url  [0]{\begingroup\@sanitize@url \@url }%
\providecommand \@url [1]{\endgroup\@href {#1}{\urlprefix }}%
\providecommand \urlprefix  [0]{URL }%
\providecommand \Eprint [0]{\href }%
\providecommand \doibase [0]{https://doi.org/}%
\providecommand \selectlanguage [0]{\@gobble}%
\providecommand \bibinfo  [0]{\@secondoftwo}%
\providecommand \bibfield  [0]{\@secondoftwo}%
\providecommand \translation [1]{[#1]}%
\providecommand \BibitemOpen [0]{}%
\providecommand \bibitemStop [0]{}%
\providecommand \bibitemNoStop [0]{.\EOS\space}%
\providecommand \EOS [0]{\spacefactor3000\relax}%
\providecommand \BibitemShut  [1]{\csname bibitem#1\endcsname}%
\let\auto@bib@innerbib\@empty
%</preamble>
\bibitem [{\citenamefont {Derrida}\ \emph {et~al.}(1995)\citenamefont
  {Derrida}, \citenamefont {Hakim},\ and\ \citenamefont
  {Pasquier}}]{PhysRevLett.75.751}%
  \BibitemOpen
  \bibfield  {author} {\bibinfo {author} {\bibfnamefont {B.}~\bibnamefont
  {Derrida}}, \bibinfo {author} {\bibfnamefont {V.}~\bibnamefont {Hakim}},\
  and\ \bibinfo {author} {\bibfnamefont {V.}~\bibnamefont {Pasquier}},\
  }\bibfield  {title} {\bibinfo {title} {Exact first-passage exponents of 1d
  domain growth: Relation to a reaction-diffusion model},\ }\href
  {https://doi.org/10.1103/PhysRevLett.75.751} {\bibfield  {journal} {\bibinfo
  {journal} {Phys. Rev. Lett.}\ }\textbf {\bibinfo {volume} {75}},\ \bibinfo
  {pages} {751} (\bibinfo {year} {1995})}\BibitemShut {NoStop}%
\bibitem [{\citenamefont {Krug}\ \emph {et~al.}(1997)\citenamefont {Krug},
  \citenamefont {Kallabis}, \citenamefont {Majumdar}, \citenamefont {Cornell},
  \citenamefont {Bray},\ and\ \citenamefont {Sire}}]{PhysRevE.56.2702}%
  \BibitemOpen
  \bibfield  {author} {\bibinfo {author} {\bibfnamefont {J.}~\bibnamefont
  {Krug}}, \bibinfo {author} {\bibfnamefont {H.}~\bibnamefont {Kallabis}},
  \bibinfo {author} {\bibfnamefont {S.~N.}\ \bibnamefont {Majumdar}}, \bibinfo
  {author} {\bibfnamefont {S.~J.}\ \bibnamefont {Cornell}}, \bibinfo {author}
  {\bibfnamefont {A.~J.}\ \bibnamefont {Bray}},\ and\ \bibinfo {author}
  {\bibfnamefont {C.}~\bibnamefont {Sire}},\ }\bibfield  {title} {\bibinfo
  {title} {Persistence exponents for fluctuating interfaces},\ }\href
  {https://doi.org/10.1103/PhysRevE.56.2702} {\bibfield  {journal} {\bibinfo
  {journal} {Phys. Rev. E}\ }\textbf {\bibinfo {volume} {56}},\ \bibinfo
  {pages} {2702} (\bibinfo {year} {1997})}\BibitemShut {NoStop}%
\bibitem [{\citenamefont {Newman}\ and\ \citenamefont
  {Stein}(1999)}]{PhysRevLett.82.3944}%
  \BibitemOpen
  \bibfield  {author} {\bibinfo {author} {\bibfnamefont {C.~M.}\ \bibnamefont
  {Newman}}\ and\ \bibinfo {author} {\bibfnamefont {D.~L.}\ \bibnamefont
  {Stein}},\ }\bibfield  {title} {\bibinfo {title} {Blocking and persistence in
  the zero-temperature dynamics of homogeneous and disordered ising models},\
  }\href {https://doi.org/10.1103/PhysRevLett.82.3944} {\bibfield  {journal}
  {\bibinfo  {journal} {Phys. Rev. Lett.}\ }\textbf {\bibinfo {volume} {82}},\
  \bibinfo {pages} {3944} (\bibinfo {year} {1999})}\BibitemShut {NoStop}%
\bibitem [{\citenamefont {Brown}\ \emph {et~al.}(1997)\citenamefont {Brown},
  \citenamefont {Dell},\ and\ \citenamefont {Wood}}]{brown1997optimization}%
  \BibitemOpen
  \bibfield  {author} {\bibinfo {author} {\bibfnamefont {G.~G.}\ \bibnamefont
  {Brown}}, \bibinfo {author} {\bibfnamefont {R.~F.}\ \bibnamefont {Dell}},\
  and\ \bibinfo {author} {\bibfnamefont {R.~K.}\ \bibnamefont {Wood}},\
  }\bibfield  {title} {\bibinfo {title} {Optimization and persistence},\
  }\href@noop {} {\bibfield  {journal} {\bibinfo  {journal} {Interfaces}\
  }\textbf {\bibinfo {volume} {27}},\ \bibinfo {pages} {15} (\bibinfo {year}
  {1997})}\BibitemShut {NoStop}%
\bibitem [{\citenamefont {Widmer}\ and\ \citenamefont
  {Kubat}(1996)}]{widmer1996learning}%
  \BibitemOpen
  \bibfield  {author} {\bibinfo {author} {\bibfnamefont {G.}~\bibnamefont
  {Widmer}}\ and\ \bibinfo {author} {\bibfnamefont {M.}~\bibnamefont {Kubat}},\
  }\bibfield  {title} {\bibinfo {title} {Learning in the presence of concept
  drift and hidden contexts},\ }\href@noop {} {\bibfield  {journal} {\bibinfo
  {journal} {Machine learning}\ }\textbf {\bibinfo {volume} {23}},\ \bibinfo
  {pages} {69} (\bibinfo {year} {1996})}\BibitemShut {NoStop}%
\bibitem [{\citenamefont {Constantin}\ and\ \citenamefont
  {Das~Sarma}(2005)}]{PhysRevE.72.051106}%
  \BibitemOpen
  \bibfield  {author} {\bibinfo {author} {\bibfnamefont {M.}~\bibnamefont
  {Constantin}}\ and\ \bibinfo {author} {\bibfnamefont {S.}~\bibnamefont
  {Das~Sarma}},\ }\bibfield  {title} {\bibinfo {title} {Volatility,
  persistence, and survival in financial markets},\ }\href
  {https://doi.org/10.1103/PhysRevE.72.051106} {\bibfield  {journal} {\bibinfo
  {journal} {Phys. Rev. E}\ }\textbf {\bibinfo {volume} {72}},\ \bibinfo
  {pages} {051106} (\bibinfo {year} {2005})}\BibitemShut {NoStop}%
\bibitem [{\citenamefont {Majumdar}(1999)}]{majumdar1999persistence}%
  \BibitemOpen
  \bibfield  {author} {\bibinfo {author} {\bibfnamefont {S.~N.}\ \bibnamefont
  {Majumdar}},\ }\bibfield  {title} {\bibinfo {title} {Persistence in
  nonequilibrium systems},\ }\href@noop {} {\bibfield  {journal} {\bibinfo
  {journal} {Current Science}\ ,\ \bibinfo {pages} {370}} (\bibinfo {year}
  {1999})}\BibitemShut {NoStop}%
\bibitem [{\citenamefont {Bassolas}\ and\ \citenamefont
  {Nicosia}(2021)}]{bassolas2021first}%
  \BibitemOpen
  \bibfield  {author} {\bibinfo {author} {\bibfnamefont {A.}~\bibnamefont
  {Bassolas}}\ and\ \bibinfo {author} {\bibfnamefont {V.}~\bibnamefont
  {Nicosia}},\ }\bibfield  {title} {\bibinfo {title} {First-passage times to
  quantify and compare structural correlations and heterogeneity in complex
  systems},\ }\href@noop {} {\bibfield  {journal} {\bibinfo  {journal}
  {Communications Physics}\ }\textbf {\bibinfo {volume} {4}},\ \bibinfo {pages}
  {76} (\bibinfo {year} {2021})}\BibitemShut {NoStop}%
\bibitem [{\citenamefont {Li}\ and\ \citenamefont
  {Kolomeisky}(2013)}]{li2013mechanisms}%
  \BibitemOpen
  \bibfield  {author} {\bibinfo {author} {\bibfnamefont {X.}~\bibnamefont
  {Li}}\ and\ \bibinfo {author} {\bibfnamefont {A.~B.}\ \bibnamefont
  {Kolomeisky}},\ }\bibfield  {title} {\bibinfo {title} {Mechanisms and
  topology determination of complex chemical and biological network systems
  from first-passage theoretical approach},\ }\href@noop {} {\bibfield
  {journal} {\bibinfo  {journal} {The Journal of chemical physics}\ }\textbf
  {\bibinfo {volume} {139}},\ \bibinfo {pages} {10B606\_1} (\bibinfo {year}
  {2013})}\BibitemShut {NoStop}%
\bibitem [{\citenamefont {Bebon}\ and\ \citenamefont
  {Schwarz}(2022)}]{bebon2022first}%
  \BibitemOpen
  \bibfield  {author} {\bibinfo {author} {\bibfnamefont {R.}~\bibnamefont
  {Bebon}}\ and\ \bibinfo {author} {\bibfnamefont {U.~S.}\ \bibnamefont
  {Schwarz}},\ }\bibfield  {title} {\bibinfo {title} {First-passage times in
  complex energy landscapes: a case study with nonmuscle myosin ii assembly},\
  }\href@noop {} {\bibfield  {journal} {\bibinfo  {journal} {New Journal of
  Physics}\ }\textbf {\bibinfo {volume} {24}},\ \bibinfo {pages} {063034}
  (\bibinfo {year} {2022})}\BibitemShut {NoStop}%
\bibitem [{\citenamefont {Bray}\ \emph {et~al.}(2013)\citenamefont {Bray},
  \citenamefont {Majumdar},\ and\ \citenamefont
  {Schehr}}]{bray2013persistence}%
  \BibitemOpen
  \bibfield  {author} {\bibinfo {author} {\bibfnamefont {A.~J.}\ \bibnamefont
  {Bray}}, \bibinfo {author} {\bibfnamefont {S.~N.}\ \bibnamefont {Majumdar}},\
  and\ \bibinfo {author} {\bibfnamefont {G.}~\bibnamefont {Schehr}},\
  }\bibfield  {title} {\bibinfo {title} {Persistence and first-passage
  properties in nonequilibrium systems},\ }\href@noop {} {\bibfield  {journal}
  {\bibinfo  {journal} {Advances in Physics}\ }\textbf {\bibinfo {volume}
  {62}},\ \bibinfo {pages} {225} (\bibinfo {year} {2013})}\BibitemShut
  {NoStop}%
\bibitem [{\citenamefont {Lancaster}\ and\ \citenamefont
  {Godoy}(2019)}]{lancaster2019persistence}%
  \BibitemOpen
  \bibfield  {author} {\bibinfo {author} {\bibfnamefont {J.~L.}\ \bibnamefont
  {Lancaster}}\ and\ \bibinfo {author} {\bibfnamefont {J.~P.}\ \bibnamefont
  {Godoy}},\ }\bibfield  {title} {\bibinfo {title} {Persistence of power-law
  correlations in nonequilibrium steady states of gapped quantum spin chains},\
  }\href@noop {} {\bibfield  {journal} {\bibinfo  {journal} {Physical Review
  Research}\ }\textbf {\bibinfo {volume} {1}},\ \bibinfo {pages} {033104}
  (\bibinfo {year} {2019})}\BibitemShut {NoStop}%
\bibitem [{\citenamefont {Jose}(2022)}]{jose2022first}%
  \BibitemOpen
  \bibfield  {author} {\bibinfo {author} {\bibfnamefont {S.}~\bibnamefont
  {Jose}},\ }\bibfield  {title} {\bibinfo {title} {First passage statistics of
  active random walks on one and two dimensional lattices},\ }\href@noop {}
  {\bibfield  {journal} {\bibinfo  {journal} {Journal of Statistical Mechanics:
  Theory and Experiment}\ }\textbf {\bibinfo {volume} {2022}},\ \bibinfo
  {pages} {113208} (\bibinfo {year} {2022})}\BibitemShut {NoStop}%
\bibitem [{\citenamefont {Salcedo-Sanz}\ \emph {et~al.}(2022)\citenamefont
  {Salcedo-Sanz}, \citenamefont {Casillas-P{\'e}rez}, \citenamefont {Del~Ser},
  \citenamefont {Casanova-Mateo}, \citenamefont {Cuadra}, \citenamefont
  {Piles},\ and\ \citenamefont {Camps-Valls}}]{salcedo2022persistence}%
  \BibitemOpen
  \bibfield  {author} {\bibinfo {author} {\bibfnamefont {S.}~\bibnamefont
  {Salcedo-Sanz}}, \bibinfo {author} {\bibfnamefont {D.}~\bibnamefont
  {Casillas-P{\'e}rez}}, \bibinfo {author} {\bibfnamefont {J.}~\bibnamefont
  {Del~Ser}}, \bibinfo {author} {\bibfnamefont {C.}~\bibnamefont
  {Casanova-Mateo}}, \bibinfo {author} {\bibfnamefont {L.}~\bibnamefont
  {Cuadra}}, \bibinfo {author} {\bibfnamefont {M.}~\bibnamefont {Piles}},\ and\
  \bibinfo {author} {\bibfnamefont {G.}~\bibnamefont {Camps-Valls}},\
  }\bibfield  {title} {\bibinfo {title} {Persistence in complex systems},\
  }\href@noop {} {\bibfield  {journal} {\bibinfo  {journal} {Physics Reports}\
  }\textbf {\bibinfo {volume} {957}},\ \bibinfo {pages} {1} (\bibinfo {year}
  {2022})}\BibitemShut {NoStop}%
\bibitem [{\citenamefont {Okubo}(1970)}]{okubo1970horizontal}%
  \BibitemOpen
  \bibfield  {author} {\bibinfo {author} {\bibfnamefont {A.}~\bibnamefont
  {Okubo}},\ }\bibfield  {title} {\bibinfo {title} {Horizontal dispersion of
  floatable particles in the vicinity of velocity singularities such as
  convergences},\ }in\ \href@noop {} {\emph {\bibinfo {booktitle} {Deep sea
  research and oceanographic abstracts}}},\ Vol.~\bibinfo {volume} {17}\
  (\bibinfo {organization} {Elsevier},\ \bibinfo {year} {1970})\ pp.\ \bibinfo
  {pages} {445--454}\BibitemShut {NoStop}%
\bibitem [{\citenamefont {Weiss}(1991)}]{weiss1991dynamics}%
  \BibitemOpen
  \bibfield  {author} {\bibinfo {author} {\bibfnamefont {J.}~\bibnamefont
  {Weiss}},\ }\bibfield  {title} {\bibinfo {title} {The dynamics of enstrophy
  transfer in two-dimensional hydrodynamics},\ }\href@noop {} {\bibfield
  {journal} {\bibinfo  {journal} {Physica D: Nonlinear Phenomena}\ }\textbf
  {\bibinfo {volume} {48}},\ \bibinfo {pages} {273} (\bibinfo {year}
  {1991})}\BibitemShut {NoStop}%
\bibitem [{\citenamefont {Kadoch}\ \emph {et~al.}(2011)\citenamefont {Kadoch},
  \citenamefont {del Castillo-Negrete}, \citenamefont {Bos},\ and\
  \citenamefont {Schneider}}]{kadoch2011lagrangian}%
  \BibitemOpen
  \bibfield  {author} {\bibinfo {author} {\bibfnamefont {B.}~\bibnamefont
  {Kadoch}}, \bibinfo {author} {\bibfnamefont {D.}~\bibnamefont {del
  Castillo-Negrete}}, \bibinfo {author} {\bibfnamefont {W.~J.}\ \bibnamefont
  {Bos}},\ and\ \bibinfo {author} {\bibfnamefont {K.}~\bibnamefont
  {Schneider}},\ }\bibfield  {title} {\bibinfo {title} {Lagrangian statistics
  and flow topology in forced two-dimensional turbulence},\ }\href@noop {}
  {\bibfield  {journal} {\bibinfo  {journal} {Physical Review E}\ }\textbf
  {\bibinfo {volume} {83}},\ \bibinfo {pages} {036314} (\bibinfo {year}
  {2011})}\BibitemShut {NoStop}%
\bibitem [{\citenamefont {Perlekar}\ \emph {et~al.}(2011)\citenamefont
  {Perlekar}, \citenamefont {Ray}, \citenamefont {Mitra},\ and\ \citenamefont
  {Pandit}}]{PhysRevLett.106.054501}%
  \BibitemOpen
  \bibfield  {author} {\bibinfo {author} {\bibfnamefont {P.}~\bibnamefont
  {Perlekar}}, \bibinfo {author} {\bibfnamefont {S.~S.}\ \bibnamefont {Ray}},
  \bibinfo {author} {\bibfnamefont {D.}~\bibnamefont {Mitra}},\ and\ \bibinfo
  {author} {\bibfnamefont {R.}~\bibnamefont {Pandit}},\ }\bibfield  {title}
  {\bibinfo {title} {Persistence problem in two-dimensional fluid turbulence},\
  }\href {https://doi.org/10.1103/PhysRevLett.106.054501} {\bibfield  {journal}
  {\bibinfo  {journal} {Phys. Rev. Lett.}\ }\textbf {\bibinfo {volume} {106}},\
  \bibinfo {pages} {054501} (\bibinfo {year} {2011})}\BibitemShut {NoStop}%
\bibitem [{\citenamefont {Wensink}\ \emph {et~al.}(2012)\citenamefont
  {Wensink}, \citenamefont {Dunkel}, \citenamefont {Heidenreich}, \citenamefont
  {Drescher}, \citenamefont {Goldstein}, \citenamefont {L{\"o}wen},\ and\
  \citenamefont {Yeomans}}]{wensink2012meso}%
  \BibitemOpen
  \bibfield  {author} {\bibinfo {author} {\bibfnamefont {H.~H.}\ \bibnamefont
  {Wensink}}, \bibinfo {author} {\bibfnamefont {J.}~\bibnamefont {Dunkel}},
  \bibinfo {author} {\bibfnamefont {S.}~\bibnamefont {Heidenreich}}, \bibinfo
  {author} {\bibfnamefont {K.}~\bibnamefont {Drescher}}, \bibinfo {author}
  {\bibfnamefont {R.~E.}\ \bibnamefont {Goldstein}}, \bibinfo {author}
  {\bibfnamefont {H.}~\bibnamefont {L{\"o}wen}},\ and\ \bibinfo {author}
  {\bibfnamefont {J.~M.}\ \bibnamefont {Yeomans}},\ }\bibfield  {title}
  {\bibinfo {title} {Meso-scale turbulence in living fluids},\ }\href@noop {}
  {\bibfield  {journal} {\bibinfo  {journal} {Proceedings of the National
  Academy of Sciences}\ }\textbf {\bibinfo {volume} {109}},\ \bibinfo {pages}
  {14308} (\bibinfo {year} {2012})}\BibitemShut {NoStop}%
\bibitem [{\citenamefont {Dunkel}\ \emph
  {et~al.}(2013{\natexlab{a}})\citenamefont {Dunkel}, \citenamefont
  {Heidenreich}, \citenamefont {B{\"a}r},\ and\ \citenamefont
  {Goldstein}}]{dunkel2013minimal}%
  \BibitemOpen
  \bibfield  {author} {\bibinfo {author} {\bibfnamefont {J.}~\bibnamefont
  {Dunkel}}, \bibinfo {author} {\bibfnamefont {S.}~\bibnamefont {Heidenreich}},
  \bibinfo {author} {\bibfnamefont {M.}~\bibnamefont {B{\"a}r}},\ and\ \bibinfo
  {author} {\bibfnamefont {R.~E.}\ \bibnamefont {Goldstein}},\ }\bibfield
  {title} {\bibinfo {title} {Minimal continuum theories of structure formation
  in dense active fluids},\ }\href@noop {} {\bibfield  {journal} {\bibinfo
  {journal} {New Journal of Physics}\ }\textbf {\bibinfo {volume} {15}},\
  \bibinfo {pages} {045016} (\bibinfo {year} {2013}{\natexlab{a}})}\BibitemShut
  {NoStop}%
\bibitem [{\citenamefont {Dunkel}\ \emph
  {et~al.}(2013{\natexlab{b}})\citenamefont {Dunkel}, \citenamefont
  {Heidenreich}, \citenamefont {Drescher}, \citenamefont {Wensink},
  \citenamefont {B\"ar},\ and\ \citenamefont
  {Goldstein}}]{PhysRevLett.110.228102}%
  \BibitemOpen
  \bibfield  {author} {\bibinfo {author} {\bibfnamefont {J.}~\bibnamefont
  {Dunkel}}, \bibinfo {author} {\bibfnamefont {S.}~\bibnamefont {Heidenreich}},
  \bibinfo {author} {\bibfnamefont {K.}~\bibnamefont {Drescher}}, \bibinfo
  {author} {\bibfnamefont {H.~H.}\ \bibnamefont {Wensink}}, \bibinfo {author}
  {\bibfnamefont {M.}~\bibnamefont {B\"ar}},\ and\ \bibinfo {author}
  {\bibfnamefont {R.~E.}\ \bibnamefont {Goldstein}},\ }\bibfield  {title}
  {\bibinfo {title} {Fluid dynamics of bacterial turbulence},\ }\href
  {https://doi.org/10.1103/PhysRevLett.110.228102} {\bibfield  {journal}
  {\bibinfo  {journal} {Phys. Rev. Lett.}\ }\textbf {\bibinfo {volume} {110}},\
  \bibinfo {pages} {228102} (\bibinfo {year} {2013}{\natexlab{b}})}\BibitemShut
  {NoStop}%
\bibitem [{\citenamefont {Toner}\ \emph {et~al.}(2005)\citenamefont {Toner},
  \citenamefont {Tu},\ and\ \citenamefont
  {Ramaswamy}}]{toner2005hydrodynamics}%
  \BibitemOpen
  \bibfield  {author} {\bibinfo {author} {\bibfnamefont {J.}~\bibnamefont
  {Toner}}, \bibinfo {author} {\bibfnamefont {Y.}~\bibnamefont {Tu}},\ and\
  \bibinfo {author} {\bibfnamefont {S.}~\bibnamefont {Ramaswamy}},\ }\bibfield
  {title} {\bibinfo {title} {Hydrodynamics and phases of flocks},\ }\href@noop
  {} {\bibfield  {journal} {\bibinfo  {journal} {Annals of Physics}\ }\textbf
  {\bibinfo {volume} {318}},\ \bibinfo {pages} {170} (\bibinfo {year}
  {2005})}\BibitemShut {NoStop}%
\bibitem [{\citenamefont {Toner}\ and\ \citenamefont
  {Tu}(1998)}]{toner1998flocks}%
  \BibitemOpen
  \bibfield  {author} {\bibinfo {author} {\bibfnamefont {J.}~\bibnamefont
  {Toner}}\ and\ \bibinfo {author} {\bibfnamefont {Y.}~\bibnamefont {Tu}},\
  }\bibfield  {title} {\bibinfo {title} {Flocks, herds, and schools: A
  quantitative theory of flocking},\ }\href@noop {} {\bibfield  {journal}
  {\bibinfo  {journal} {Physical review E}\ }\textbf {\bibinfo {volume} {58}},\
  \bibinfo {pages} {4828} (\bibinfo {year} {1998})}\BibitemShut {NoStop}%
\bibitem [{\citenamefont {Canuto}\ \emph {et~al.}(2012)\citenamefont {Canuto},
  \citenamefont {Hussaini}, \citenamefont {Quarteroni}, \citenamefont
  {Thomas~Jr} \emph {et~al.}}]{canuto2012spectral}%
  \BibitemOpen
  \bibfield  {author} {\bibinfo {author} {\bibfnamefont {C.}~\bibnamefont
  {Canuto}}, \bibinfo {author} {\bibfnamefont {M.~Y.}\ \bibnamefont
  {Hussaini}}, \bibinfo {author} {\bibfnamefont {A.}~\bibnamefont
  {Quarteroni}}, \bibinfo {author} {\bibfnamefont {A.}~\bibnamefont
  {Thomas~Jr}}, \emph {et~al.},\ }\href@noop {} {\emph {\bibinfo {title}
  {Spectral methods in fluid dynamics}}}\ (\bibinfo  {publisher} {Springer
  Science \& Business Media},\ \bibinfo {year} {2012})\BibitemShut {NoStop}%
\end{thebibliography}%

\end{document}